\documentclass[conference]{IEEEtran}
\makeatletter
\def\ps@headings{%
\def\@oddhead{\mbox{}\scriptsize\rightmark \hfil \thepage}%
\def\@evenhead{\scriptsize\thepage \hfil \leftmark\mbox{}}%
\def\@oddfoot{}%
\def\@evenfoot{}}
\makeatother 

\IEEEoverridecommandlockouts
\usepackage{bbm}
\usepackage{amsfonts}
\usepackage[dvips]{graphicx}
\usepackage{times}
\usepackage{cite}
\usepackage{amsmath}
\usepackage{array}
\usepackage{amssymb}

\usepackage{stfloats}
\usepackage{slashbox}
\usepackage{graphicx}
\usepackage{footnote}
\usepackage{booktabs}
\usepackage{array}
\usepackage{algorithm}
\usepackage{subeqnarray}
\usepackage{cases}
\usepackage{threeparttable}
\usepackage{color}
\usepackage{hyperref}
\usepackage{epstopdf}
\usepackage{algpseudocode}
\usepackage{bm}
\usepackage{multirow}
\usepackage[labelformat=simple]{subcaption}
\usepackage{adjustbox}

\begin{document}

\title{Joint Design of Hybrid Beamforming and Phase Shifts in RIS-Aided mmWave Communication Systems}

\author{Bei Guo, Renwang Li and Meixia Tao\\
Department of Electronic Engineering, Shanghai Jiao Tong University, Shanghai, China\\
Email: \{guobei132, renwanglee, mxtao\}@sjtu.edu.cn
\thanks{This work is supported by the NSF of China under grant 61941106.}}
\maketitle
\vspace{-2.5cm}
\begin{abstract}
This paper considers a reconfigurable intelligent surface (RIS)-aided millimeter wave (mmWave) downlink communication system where hybrid analog-digital beamforming is employed at the base station (BS). We formulate a power minimization problem by jointly optimizing hybrid beamforming at the BS and the response matrix at the RIS, under signal-to-interference-plus-noise ratio (SINR) constraints. The problem is highly challenging due to the non-convex SINR constraints as well as the non-convex unit-modulus constraints for both the phase shifts at the RIS and the analog beamforming at the BS. A penalty-based algorithm in conjunction with the manifold optimization technique is proposed to handle the problem, followed by an individual optimization method with much lower complexity. 
Simulation results show that the proposed algorithm outperforms the state-of-art algorithm. Results also show that the joint optimization of RIS response matrix and BS hybrid beamforming is much superior to individual optimization.
\end{abstract}
\vspace{-0.2cm}
\section{Introduction}
\vspace{-0.3cm}
Reconfigurable Intelligent Surfaces (RISs) have emerged as a new technique to enhance wireless communications by manipulating the radio propagation environment. An RIS is an artificial meta-surface consisting of a large number of passive reflection elements that can be programmed to control the phase of the incident electromagnetic waves \cite{book}. It is appealing for communications as it can create passive beamforming (BF) towards desired receivers without radio frequency (RF) components. Compared to traditional active multi-input multi-output (MIMO) relaying, RISs are more cost-effective and do not cause any noticeable processing delay.

RISs bring a new degree of freedom to the optimization of BF design. The work \cite{8811733} studies the joint optimization of active and passive BF in an RIS-aided multi-user system for transmit power minimization under signal-to-interference-plus-noise ratio (SINR) constraints. In \cite{guo2019weighted},  the joint optimization of active and passive BF is investigated for weighted-sum-rate (WSR) maximization under transmit power constraints. The work \cite{di2019hybrid} considers the sum-rate maximization problem when only a limited number of discrete phase shifts can be realized by the RIS. Note that in all these works on joint active-passive BF design, the active BF part is fully digital as in most of the MIMO BF literature, which requires each antenna to be connected to one RF chain.

The millimeter wave (mmWave) communication over 30-300 GHz spectrum is a key technology in 5G networks to provide high data-rate transmission. A fundamental issue of mmWave communications is its sensitivity to signal blockages due to the high frequency band. Thus, an important use case of RISs is to overcome the blockage effect in mmWave systems. Compared with the sub-6 GHz systems, mmWave systems suffer much higher hardware cost and power consumption on the RF circuits. Hybrid analog and digital (A/D) BF is more favorable than fully digital BF since it allows multiple antennas to share one RF chain \cite{8030501}. It is therefore desirable to consider hybrid BF for the active BF design in RIS-aided mmWave communications. Recently, the work \cite{xiu2020reconfigurable} focuses on WSR maximization in a nonorthogonal multiple access system by jointly designing the RIS phase shifts and hybrid BF. The work \cite{ying2020gmdbased} proposes an individual design algorithm for the hybrid  beamformer, and the RIS response matrix to achieve low error rate in a wideband system.
\begin{figure}
\begin{centering}
\includegraphics[scale=.33]{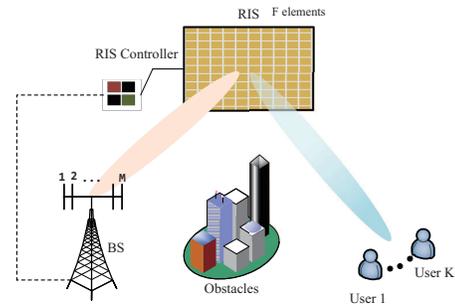}
\vspace{-0.3cm}
 \caption{\small{RIS-aided downlink mmWave communication system.}}\label{fig_systemmodel}
\end{centering}
\vspace{-0.8cm}
\end{figure}

 This paper considers the joint optimization of the hybrid BF at the BS and the phase shifts at the RIS in an RIS-aided multi-user mmWave system. We formulate an optimization problem for minimizing the total transmit power at the BS subject to individual SINR constraints for each user. This problem is highly non-convex and very challenging due to two main obstacles. One is that all variables are tightly coupled in the constraints. To tackle this issue, we reformulate the problem using the penalty function method. More specifically, we introduce auxiliary variables to decouple these variables and then add the associated equality constraints to the objective function as penalty terms. Another obstacle is that both the phase shifts at the RIS and the analog beamformers at the BS have unit-modulus constraints. Unlike the conventional semidefinite relaxation (SDR) method  \cite{8811733}, we adopt a manifold optimization technique to handle these unit-modulus constraints. Overall, we propose a two-layer penalty-based algorithm in conjunction with the Riemannian manifold optimization to find a stationary solution to the original problem. Simulation results show that the proposed penalty-based algorithm outperforms the traditional SDR-based optimization algorithm. Results also show that the proposed hybrid beamforming at the BS can perform closely to the fully digital beamforming. 
\vspace{-0.3cm}
\section{System Model And Problem Formulation}
\vspace{-0.2cm}
\subsection{System Model}
\vspace{-0.2cm}
As shown in Fig.~\ref{fig_systemmodel}, we consider an RIS-aided downlink mmWave system where one BS, equipped with $M$ antennas, communicates with $K$ single-antenna users via the help of one RIS with $F$ unit cells. The BS employs the sub-connected hybrid A/D beamforming structure with $N$ RF chains, each connected to $D=M/N$ antennas. Let $s_j$ denote the information signal intended to user $j$, for $j\in \mathcal{K} \triangleq \{1,\ldots,K\}$. It is assumed to be independent to each other and satisfies $E[|s_j|^2]=1$. Each of these signals is first weighted by a digital beamforming vector, denoted as $\mathbf{w}_j\in\mathbb{C}^{N\times{1}}$. These weighted signal vectors are summed together and each entry is sent to an RF chain, then multiplied by an analog beamforming vector, denoted as $\mathbf{v}_n\in \mathbb{C}^{D\times{1}}$, for $n\in \mathcal{N}\triangleq \{1,2,\cdots,N\}$. Each entry of $\mathbf{v}_n$, denoted as ${v}_{n,d}, \forall d\in \mathcal{D}\triangleq\{1,\ldots,D\}$ is a phase shifter, i.e., $|v_{n,d}|=1$. The overall analog beamforming matrix can be represented as $\mathbf{V}=\text{diag}\{\mathbf{v}_1,\cdots,\mathbf{v}_N\} \in \mathbb{C}^{M\times N}$. At the RIS, let $\mathcal{F} \triangleq \{1,2,\cdots,F\}$ denote the set of total RIS unit cells, and define a diagonal matrix $\mathbf{\Theta}= \text{diag} {(b_1,b_2,\ldots,b_F)}$ as the response-coefficient matrix, where $b_f=e^{j\theta_f}, \theta_f\in{[0,2\pi)}$ being the phase shift of the $f$th unit cell. The total transmit power of the BS is given by
\vspace{-0.1cm}
\begin{equation}
    P_\text{total}=\sum\limits_{k=1}^{K}\Arrowvert{\mathbf{V}\mathbf{w}_k}\Arrowvert^2
     =D \sum\limits_{k=1}^K  \Arrowvert{\mathbf{w}_k}\Arrowvert^2.
\end{equation}
\vspace{-0.1cm}
We assume the BS-user link is blocked, and thus the direct path can be ignored. The channel state information (CSI) of all links is assumed to be perfectly known at the BS and all the channels experience quasi-static flat-fading.

The received signal of user $k$ can be represented as
\begin{equation}
y_k=\mathbf{h}_{k}^{H}\mathbf{\Theta}\mathbf{G}\mathbf{V}\sum\limits_{j=1}^{K}{\mathbf{w}_js_j}+n_k, \forall{k}\in\mathcal{K},
\end{equation}
where  $\mathbf{G}\in \mathbb{C}^{F \times M}$ is the channel matrix from the BS to the RIS, $\mathbf{h}_{k}^{H} \in\mathbb{C}^{1\times{F}}$ is the channel vector from the RIS to user k, and $n_k\thicksim\mathcal{C}\mathcal{N}{(0, \sigma_k^2)}$ is the additive white Gaussian noise at the receiver of user $k$.

The received SINR of user $k$ can be expressed as
\begin{equation}
\text{SINR}_k=\frac{\arrowvert{\mathbf{h}_{k}^H\mathbf{\Theta}\mathbf{G}\mathbf{V}\mathbf{w}_k} \arrowvert^2}{\sum _{j\not=k}\arrowvert{\mathbf{h}_{k}^H\mathbf{\Theta}\mathbf{G} \mathbf{V}\mathbf{w}_j}\arrowvert^2+\sigma_k^2}, \forall{k}\in\mathcal{K}.
\end{equation}
\subsection{mmWave Channel Model}
We adopt the widely used Saleh-Valenzuela channel model \cite{6717211} for mmWave communications. Specifically, the channel matrix between BS and RIS can be written as
\begin{equation}
\small{\mathbf{G}=\small{\sqrt{\frac{MF}{{N_\text{cl}}_1{N_\text{ray}}_1}} \sum \limits_{i_1=1}^{{N_\text{cl}}_1} \sum\limits_{l_1=1}^{{N_\text{ray}}_1} \alpha_{i_1l_1}\mathbf{a}_R(\phi_{i_1l_1}^{Rr}, \delta_{i_1l_1}^{Rr})\mathbf{a}_B(\phi_{i_1l_1}^{B}, \delta_{i_1l_1}^{B})^H}}.
\vspace{-0.1cm}
\end{equation}
Here, ${N_\text{cl}}_1$ denotes the number of scattering clusters,  ${N_\text{ray}}_1$ denotes the number of rays in each cluster, $\alpha_{i_1l_1}$ denotes the channel coefficient of the $l_1$th ray in the $i_1$th propagation cluster. Moreover, $\mathbf{a}_R(\phi_{i_1l_1}^{Rr},\delta_{i_1l_1}^{Rr})$ and $\mathbf{a}_B(\phi_{i_1l_1}^B,\delta_{i_1l_1}^B)$ represent the receive array response vectors of the RIS and the transmit array response vectors of the BS respectively, where $\phi_{i_1l_1}^{Rr}(\phi_{i_1l_1}^{B})$ and $\delta_{i_1l_1}^{Rr}(\delta_{i_1l_1}^B)$ represent azimuth and elevation angles of arriving at the RIS (or departing from the BS).

The channel vector between the RIS and the $k$-th user can be represented as
\begin{equation}
\mathbf{h}_{k}^H=\sqrt{\frac{F}{{N_\text{cl}}_2{N_\text{ray}}_2}} \sum \limits_{i_2=1}^{{N_\text{cl}}_2} \sum\limits_{l_2=1}^{{N_\text{ray}}_2} \beta_{i_2l_2}\mathbf{a}_R(\phi_{i_2l_2}^{Rt},\delta_{i_2l_2}^{Rt})^H.
\end{equation}
Here, ${N_\text{cl}}_2$, ${N_\text{ray}}_2$, $\beta_{i_2l_2}$, $\phi_{i_2l_2}^{Rt}$ and $\delta_{i_2l_2}^{Rt}$ are defined in the same way as above.

We consider the uniform planar array (UPA) structure at both BS and RIS. The array response vector can be denoted as
 \begin{equation}
    \begin{aligned}
      \label{upa}
        \mathbf{a}_{{z}}\left(\phi, \delta\right)= \frac{1}{\sqrt{A_1 A_2}}\left[1, \ldots, e^{j \frac{2 \pi}{\lambda} d_1 \left(o \sin \phi \sin \delta+p \cos \delta\right)}\right.\\
        \left.\ldots, e^{j \frac{2 \pi}{\lambda} d_1 \left((A_1-1) \sin \phi \sin \delta)+(A_2-1) \cos \delta\right)}\right]^{T},
    \end{aligned}
  \end{equation}
where ${z} \in \{R,B\}$, $\lambda$ is the signal wavelength, $d_1$ is the antenna or unit cell spacing which is assumed to be half wavelength distance, $0\leq{o}<A_1$ and $0\leq{p}<A_2$, $A_1$ and $A_2$ represent the number of rows and columns of the UPA in the 2D plane, respectively.
\subsection{Problem Formulation}
\label{pro_formulation}
We aim to minimize the transmit power by jointly optimizing the digital beamforming matrix  $\mathbf{W}=\left[\mathbf{w}_1, \cdots,\mathbf{w}_K \right]\in \mathbb{C}^{N\times K}$ and the analog beamforming matrix $\mathbf{V}$ at the BS, as well as the overall response-coefficient matrix $\mathbf{\Theta}$ at the RIS, subject to a minimum SINR constraint for each user. Thus, the optimization problem can be formulated as
    \begin{subequations}\label{prob_original}
    \begin{align}
            &{\mathcal{P}_0:}&{\min \limits_{\{\mathbf{V},\mathbf{W},\mathbf{\Theta}\}}} \quad & {D\sum \limits_{k=1}^{K}\left\| \mathbf{w}_{k}\right\|^{2}} \\
            & &{\text { s.t. }}\quad & {\text{SINR}_{k} \geq \gamma_{k}, \forall k\in\mathcal{K}},  \label{const1}\\
            & &{}&{\left| v_{n,d} \right|=1, \forall n \in \mathcal{N}, \forall d \in \mathcal{D}}, \label{const2}\\
            & &{}&{|b_f|=1, \forall f \in \mathcal{F}}, \label{const3}
    \end{align}
    \end{subequations}
where $\gamma_k>0$ is the minimum SINR requirement of user $k$.

The problem is non-convex due to the non-convex SINR constraints \eqref{const1} and the unit-modulus constraints \eqref{const2}, \eqref{const3}. A commonly used approach to solve this type of optimization problems approximately is to apply the block coordinate descent (BCD) techniques in conjunction with the SDR method. More specifically, the digital beamforming matrix $\mathbf{W}$, the analog beamforming matrix $\mathbf{V}$, and the RIS  response-coefficient matrix $\mathbf{\Theta}$ are updated in an alternating manner in each iteration. The sub-problem of finding $\mathbf{W}$ can be solved by second-order cone program (SOCP) method, and both the sub-problems of finding $\mathbf{V}$ and finding $\mathbf{\Theta}$ can be solved by SDR. However, the solution obtained by SDR is not guaranteed to be rank-one and additional randomization approach is needed. In addition, when the number of users is close to the number of RF chains at the BS, the randomization procedure may fail to find a feasible solution.
\vspace{-0.1cm}
\section{Penalty-based Joint Optimization Algorithm}
In this section, we propose a two-layer penalty-based algorithm for the considered problem $\mathcal{P}_0$. The BCD method is adopted in the inner layer to solve a penalized problem and the penalty factor is updated in the outer layer until converge. Specifically, we firstly introduce auxiliary variables $\{t_{k,j}\}$ to represent $ \mathbf{h}_{k}^{H} \mathbf{\Theta} \mathbf{G} \mathbf{V} \mathbf{w}_j$ such that variables $\mathbf{W}$,$\mathbf{V}$ and $\mathbf{\Theta}$ can be decoupled. Then, the non-convex constraints \eqref{const1} can be equivalently written as
\begin{subequations}
    \begin{align}
        {\frac{ \left| t_{k,k} \right|^{2}} {\sum_{j \neq k}^{K}\left| t_{k,j} \right|^{2}+\sigma_{k}^{2}} \geq \gamma_{k}, \forall k \in \mathcal{K},}\label{penalty_ori_const1}\\
		{t_{k,j}= \mathbf{h}_{k}^{H} \mathbf{\Theta} \mathbf{G} \mathbf{V} \mathbf{w}_j}, \forall k,j \in \mathcal{K}. \label{penalty_ori_const2}
    \end{align}
\end{subequations}
Then, the equality constraints \eqref{penalty_ori_const2} is relaxed and added to the objective function as a penalty term. Thereby, the original problem $\mathcal{P}_0$ can be converted to
 \begin{equation}\label{penalty_prob}
    \begin{array}{lll}
            \hspace{-0.5cm}{\mathcal{P}_1:}& \hspace{-0.4cm}{\min \limits_{\small{\mathbf{V}, \mathbf{W}, \mathbf{\Theta},\{t_{k,j}\}}}} & \hspace{-0.4cm} {\small{D \sum \limits_{k=1}^{K}\left\| \mathbf{w}_{k}\right\|^{2} +
            \frac{\rho}{2} \sum \limits_{j=1}^K \sum \limits_{k=1}^K \left| \mathbf{h}_{k}^{H} \mathbf{\Theta} \mathbf{G} \mathbf{V} \mathbf{w}_j-t_{k,j} \right|^2}} \\
            & {\text { s.t. }} & {\eqref{penalty_ori_const1},\eqref{const2},\eqref{const3} },
    \end{array}
\end{equation}
where $\rho>0$ is the penalty factor. The choice of $\rho$ is crucial to balance the original objective function and the equality constraints. It is seen that the objective function in $\mathcal{P}_1$ is dominated by the penalty term when $\rho$ is large enough and consequently, the equality constraints \eqref{penalty_ori_const2} can be well met by the solution. Thus, we can start with a small value of $\rho$ to get a good start point, and then by gradually increasing $\rho$, a high precision solution can be obtained. Similar approach is adopted in \cite{wu2019joint}.

\subsection{Inner Layer: BCD Algorithm for Solving Problem $\mathcal{P}_1$}
For any given $\rho$, the problem $\mathcal{P}_1$ is non-convex but all the optimization variables $\{\mathbf{V}, \mathbf{W}, \mathbf{\Theta}, \{t_{k,j}\}\}$ are decoupled in the constraints. We therefore adopt BCD method to optimize each of them alternately while keeping the others fixed.
\subsubsection{Optimize $\mathbf{W}$}
When $\mathbf{V}$,$\mathbf{\Theta}$ and $\{t_{k,j}\}$ are fixed, problem $\mathcal{P}_1$ becomes a non-constraint convex optimization problem. Thus, the optimal $\mathbf{W}$ can be obtained by the first-order optimality condition, i.e.,
\begin{equation}\label{penalty_optimw}
    \mathbf{w}_k=\rho \mathbf{A}^{-1} \sum \limits_{j=1}^K \tilde{\mathbf{h}}_j^H t_{j,k}, \forall k \in \mathcal{K},
\end{equation}
where $ \tilde{\mathbf{h}}_j=\mathbf{h}_{ j}^{H} \mathbf{\Theta} \mathbf{G} \mathbf{V}$ and $\mathbf{A}=2D \mathbf{I}_N + \rho \sum \limits_{j=1}^K \tilde{\mathbf{h}}_j^H \tilde{\mathbf{h}}_j$.

\subsubsection{Optimize $\mathbf{\Theta}$}
\label{manifold_b}
Let $\mathbf{b}=[b_1,b_2,\ldots,b_F]^{H}$. When other variables are fixed, problem $\mathcal{P}_1$ is reduced to (with constant terms ignored)
\begin{subequations} \label{penalty_optimb}
    \begin{align}
        {\min \limits_{\mathbf{b}}} \quad & { f(\mathbf{b})=\sum \limits_{j=1}^K \sum \limits_{k=1}^K \left| \mathbf{b}^H \mathbf{c}_{k,j} -t_{k,j} \right|^2} \label{penalty_obtimb_obj}\\
        {\text { s.t. }} \quad & { |b_f|=1, \forall f \in \mathcal{F}}\label{unitm},
    \end{align}
\end{subequations}
where $\mathbf{c}_{k,j}= \text{diag} (\mathbf{h}_{k}^H) \mathbf{G} \mathbf{V} \mathbf{w}_j \in \mathbb{C}^{F\times1}$. Although the objective function is convex for $\mathbf{b}$, the problem \eqref{penalty_optimb} is still non-convex due to the unit-modulus constraints \eqref{unitm}. To handle this problem, one way is to alternately optimize the $F$ units one by one as in \cite{guo2019weighted,wu2019joint}. Although closed-form expression is available for each unit, this method is still inefficient since the unit number $F$ is usually very large. Another way is to adopt the SDR technique as in \cite{8811733}. But its complexity is high and additional randomization procedure is needed. Note that the unit-modulus constraints \eqref{unitm} form a complex circle manifold $\mathcal{M}= \{\mathbf{b}\in \mathbb{C}^F: |b_1|=\cdots=|b_F|\}$\cite{absil2009optimization}. Therefore, different from the above approaches, we adopt the manifold optimization technique to solve this problem efficiently and optimally. In specific, we adopt the Riemannian conjugate gradient (RCG) algorithm. The RCG algorithm is widely applied in hybrid beamforming design \cite{yu2016alternating} and recently applied in RIS-aided systems as well\cite{yu2019miso},\cite{guo2020weighted}. Each iteration of the RCG algorithm involves three key steps, namely, to compute Riemannian gradient, to find search direction and retraction.

The Riemannian gradient $\operatorname{grad}_\mathbf{b} f(\mathbf{b})$ of the function $f(\mathbf{b})$ is defined as the orthogonal projection of the Euclidean gradient $\nabla f(\mathbf{b})$ onto the tangent space ${T}_\mathbf{b} \mathcal{M}$ of the manifold $\mathcal{M}$, which can be expressed as
\begin{equation}
T_{\mathbf{b}} \mathcal{M}=\left\{\mathbf{z} \in \mathbb{C}^{M}: \Re\left\{\mathbf{z} \odot \mathbf{b}^{*}\right\}=\mathbf{0}_{M}\right\},
\end{equation}
where $\odot$ denotes the Hadamard product. The Euclidean gradient of $f(\mathbf{b})$ over $\mathbf{b}$ is given by
\begin{equation}
    \nabla f(\mathbf{b}) = 2\sum \limits_{j=1}^K \sum \limits_{k=1}^K \mathbf{c}_{k,j} (\mathbf{c}_{k,j}^H \mathbf{b} - t_{k,j}^H).
\end{equation}
Then, the Riemannian gradient is given by
\begin{equation}
\operatorname{grad}_{\mathbf{b}} f(\mathbf{b})=\nabla f(\mathbf{b}) -\operatorname{Re}\left\{\nabla f(\mathbf{b}) \odot \mathbf{b}\right\} \odot \mathbf{b}.
\end{equation}

With the Riemannian gradient, we can update the search direction $\mathbf{d}$ by conjugate gradient method, i.e.,
\begin{equation}
\mathbf{d}=-\operatorname{grad} f_{\mathbf{b}}+\lambda_1 \mathcal{T}(\overline{\mathbf{d}}),
\end{equation}
where $\lambda_1$ is the update parameter, $\overline{\mathbf{d}}$ is the previous search direction and $\mathcal{T}(\mathbf{d})=\overline{\mathbf{d}}-\operatorname{Re} \left\{\overline{\mathbf{d}} \odot \mathbf{b}^{*}\right\} \odot \mathbf{b}$.

Since the updated point may leave the previous manifold space, a retraction operation $\operatorname{Retr}_{\mathbf{b}}$ is needed to project the point to the manifold itself:
\vspace{-0.1cm}
\begin{equation}
\operatorname{Retr}_{\mathbf{b}}: \small{\mathbf{b}_{f} \leftarrow \frac{\left(\mathbf{b}+\lambda_{2} \mathbf{d}\right)_{f}}{\left|\left(\mathbf{b}+\lambda_{2} \mathbf{d}\right)_{f}\right|}},
\vspace{-0.1cm}
\end{equation}
where $\lambda_2$ is the Armijo backtracking line search step size.

\subsubsection{Optimize $\mathbf{V}$}
Define
    $
     \mathbf{x} \triangleq\left[\mathbf{v}_{1}^{T}, \mathbf{v}_{2}^{T}, \ldots, \mathbf{v}_{N}^{T}\right]^{T} \in \mathbb{C}^{M\times 1},
    $
    and
    $
    \mathbf{Z}_j \triangleq \text{diag}\{{w}_{ j,1}\mathbf{I}_D, \ldots, {w}_{ j,N}\mathbf{I}_D \} \in \mathbb{C}^{M\times M},
    $
 where $\left|{x}_m\right| = 1, \forall m\in \mathcal{M}\triangleq \{1,2,\cdots,M\}$ and ${w}_{j,n}$ denotes the $n$-th entry of $\mathbf{w}_j$. Then, we have
    $
        \mathbf{V} \mathbf{w}_j=\mathbf{Z}_j \mathbf{x} \in \mathbb{C}^{M\times 1}.
    $
When other variables are fixed, problem $\mathcal{P}_1$ is given by
\begin{subequations} \label{penalty_optimx}
    \begin{align}
        {\min \limits_{\mathbf{x}}}\quad & { f(\mathbf{x})=\sum \limits_{j=1}^K \sum \limits_{k=1}^K \left| \mathbf{d}_{k,j} \mathbf{x} -t_{k,j} \right|^2} \\
        {\text { s.t. }}\quad & { |{x}(m)|=1, \forall m \in \mathcal{M}},
    \end{align}
\end{subequations}
where $\mathbf{d}_{k,j}= \mathbf{b}^H \text{diag} (\mathbf{h}_{k}^H) \mathbf{G} \mathbf{Z}_j \in \mathbb{C}^{1\times M}$. Similar to Section \ref{manifold_b}, it can be effectively solved by the RCG algorithm and the details are skipped.

\subsubsection{Optimize $\{t_{k,j}\}$}
With other variables fixed, problem $\mathcal{P}_1$ can be reduced to
\begin{subequations} \label{penalty_t}
    \begin{align}
        {\min \limits_{\{t_{k,j}\}}} \quad & { \sum \limits_{j=1}^K \sum \limits_{k=1}^K \left| \mathbf{h}_{k}^{H} \mathbf{\Theta} \mathbf{G} \mathbf{V} \mathbf{w}_j-t_{k,j} \right|^2} \\
        {\text { s.t. }} \quad & {\frac{ \left| t_{k,k} \right|^{2}} {\sum_{j \neq k}^{K}\left| t_{k,j} \right|^{2}+\sigma_{k}^{2}} \geq \gamma_{k}, \forall k \in \mathcal{K}}\label{22}.
    \end{align}
\end{subequations}

The objective function is convex over $\{t_{k,j}\}$. Although constraints \eqref{22} are still non-convex, they can be translated to the form of second-order cone, which can be effectively and optimally solved by SOCP method \cite{1561584}.

\subsection{Outer Layer: Update Penalty factor}
The penalty factor $\rho$ is initialized to be a small number to find a good start point, then gradually increased to tighten the penalty. Specifically,
\begin{equation}\label{penalty_rho}
    \rho:=\frac{\rho}{c}, 0 <c <1,
\end{equation}
where $c$ is a scaling parameter. A larger $c$ may lead to a more precise solution with higher running time.

\subsection{Algorithm}
 \begin{algorithm}[t]
    \caption{Penalty-based Algorithm with Manifold Optimization}
    \label{alg_penalty}
    \begin{algorithmic}[1]
    \State Initialize $\mathbf{V}$, $\mathbf{\Theta}$, $\rho$ and $\{t_{k,j}\},\forall k,j \in \mathcal{K}$.
    \Repeat
        \Repeat
            \State Update $\mathbf{W}$ by \eqref{penalty_optimw};
            \State Update $\mathbf{\Theta}$ by solving problem \eqref{penalty_optimb};
            \State Update $\mathbf{V}$ by solving problem \eqref{penalty_optimx};
            \State Update $\{t_{k,j}\}$ by solving problem \eqref{penalty_t};
        \Until The decrease of the objective value of problem $\mathcal{P}_1$ is below threshold $\epsilon_1>0$.
        \State Update $\rho$ by \eqref{penalty_rho}.
    \Until The stopping indicator $\xi$ is below threshold $\epsilon_2>0$.
    \end{algorithmic}
 \end{algorithm}
The overall penalty-based algorithm is summarized in \textit{Algorithm} \ref{alg_penalty}. Define the stopping indicator $\xi$ as following
\begin{equation}\label{stop_criteria}
\vspace{0.2cm}
\xi \triangleq \max \left\{ | \mathbf{h}_{k}^{H} \mathbf{\Theta} \mathbf{G} \mathbf{V} \mathbf{w}_j-t_{k,j} |^2, \forall k,j \in \mathcal{K} \right\}.
\vspace{-0.3cm}
\end{equation}
When $\xi$ is below a pre-defined threshold $\epsilon_2>0$, the equality constraints \eqref{penalty_ori_const2} are considered to be satisfied and the proposed algorithm is terminated. Since we start with a small penalty and gradually increase its value, the objective value of problem $\mathcal{P}_1$ is finally determined by the penalty part and the equality constraints are guaranteed to be satisfied. Note that, for any given $\rho$, problem $\mathcal{P}_1$ is solved through the BCD method and each subproblem can obtain an optimal solution. Thus, \textit{Algorithm} \ref{alg_penalty} is guaranteed to converge to a stationary point. The total complexity of \textit{Algorithm} \ref{alg_penalty} is $\mathcal{O}(I_OI_J(KN^3+I_QK^2F+I_VK^2M+K^7))$, where $I_O$, $I_J$, $I_Q$, and $I_V$ denote the iteration times of the outer loop, the inner loop, the inner RCG algorithm to update $\mathbf{\Theta}$, and the inner RCG algorithm to update $\mathbf{\mathbf{V}}$, respectively.

\section{Individual Optimization}
\label{individual}
To reduce the complexity of solving problem $\mathcal{P}_0$, we develop an individual optimization approach in this section, where the RIS response matrix $\mathbf{\Theta}$, the analog beamformer $\mathbf{V}$, and the digital beamformer $\mathbf{W}$ are obtained sequentially without alternating optimization.

\subsection{RIS design}
The equivalent channel between the BS and the $k$th user via the RIS can be represented as $\mathbf{h}_{k}^H  \mathbf{\Theta} \mathbf{G}$. To ensure the receive signal quality of each user, we aim to find the optimal RIS response matrix for maximizing the equivalent channel gain of the user who has the worst channel state, i.e.,\begin{subequations}\label{max_min_ris}
        \begin{align}
            {\max \limits_{\mathbf{\Theta}}}\quad &{\min \limits_{k\in \mathcal{K}} \| \mathbf{h}_{k}^H \mathbf{\Theta}\mathbf{G} \|^2} \\
            {\text { s.t. }}\quad & {|b_f|=1, \forall f \in \mathcal{F}.}
        \end{align}
        \vspace{-0.2cm}
\end{subequations}
This problem can be effectively solved by SDR.
\subsection{Analog BF design}
Orthogonal match pursuit (OMP) method is widely adopted to design the analog beamformer\cite{6717211}. If the BS adopts the fully digital BF structure, the optimal digital BF under the zero-forcing (ZF) scheme is given by $\mathbf{F}_{\text {opt}}=\tilde{\mathbf{H}}^\dagger \text{diag}(\sqrt{\gamma_1 \sigma_1^2}, \ldots, \sqrt{\gamma_K \sigma_K^2})$, where $\tilde{\mathbf{H}}=\left[(\mathbf{h}_1^H \mathbf{\Theta} \mathbf{G})^T, \ldots, (\mathbf{h}_K^H \mathbf{\Theta} \mathbf{G})^T \right]^T$ and $\dagger$ denotes the pseudo-inverse. Define the overlapping coefficient as $\mu$, and denote the codebook as $\mathbf{A}=[\mathbf{a}_B (\psi_1, \phi_1), \ldots, \mathbf{a}_B (\psi_1, \phi_{\mu N_z}), \ldots, \mathbf{a}_B (\psi_{\mu N_y}, \phi_{\mu N_z})]$, where $N_y$ and $N_z$ denote the horizontal and vertical length, $\psi_i= \frac{2 \pi(i-1)}{\mu N_{y}}, i=1,2,\ldots,\mu N_y$ and
$\phi_j =\frac{2 \pi(j-1)}{\mu N_{z}}, j=1,2,\ldots,\mu N_z$, respectively.
Then, we can use a selection matrix $\mathbf{T}\in \mathbb{R}^{\mu^2 N_y N_z \times N}$ to select proper columns. Specifically, the analog BF problem can be formulated as
\begin{subequations} \label{analog_codebook}
    \begin{align}
        {\mathbf{T}^*=\underset{\mathbf{T}}{\arg \min } } \quad & {  \left\|\mathbf{F}_{\text {opt}}- \mathbf{A}_t \mathbf{T} \mathbf{F}_{BB} \right\|_{F} }\\
         {\text{ s.t. }}\quad & {\left\|\operatorname{diag}\left(\mathbf{T} \mathbf{T}^{H}\right)\right\|_{0}= N, }\\
         {}& {\mathbf{A}_t= \mathbf{I}_{t} \odot \mathbf{A}, t\in \mathcal{N},}
    \end{align}
\end{subequations}
where $\mathbf{I}_t$ is a $M\times1$ zero-vector with the entry from $(t-1)D+1$ to $tD$ being one; $\odot$ denotes the Hadamard product. Since the structure of analog BF is sub-connected, we use $\mathbf{I}_t$ to modify the codebook. Then, the OMP method can be applied to obtain the optimal $\mathbf{T}^*$. The analog BF can be recovered, i.e., $\mathbf{V}=\mathbf{A}_t \mathbf{T}^*$.
\vspace{-0.2cm}
\subsection{Digital BF design}
After obtaining the RIS phase shifts and the analog beamformer, we need to obtain the optimal digital BF vector by solving following problem,
\vspace{-0.1cm}
 \begin{equation}\label{inde_w}
    \begin{array}{ll}
            {\min \limits_{ \mathbf{W}}}\quad & {D \sum \limits_{k=1}^{K}\left\| \mathbf{w}_{k}\right\|^{2}} \\
            {\text { s.t. }} & {\eqref{const1} }.
    \end{array}
    \vspace{-0.2cm}
\end{equation}

Note that it is the conventional power minimization problem in the multi-input single-output (MISO) downlink system, which can be effectively solved by SOCP method.

\section{Simulation Results}
We consider a $6\times 6$ UPA structure at the BS with a total of $M=36$ antennas and $N=6$ RF chains located at (0 m, 0 m). The RIS is located at ($d_{RIS}$ m, 10 m) and equipped with $F_1 \times F_2$ unit cells where $F_1=6$ and $F_2$ can vary. Users are uniformly and randomly distributed in a circle centered at (100 m, 0 m) with radius 5 m. As for the channel, we set ${N_{\text{cl}}}_1={N_\text{cl}}_2=2$ clusters, ${N_\text{ray}}_1={N_\text{ray}}_2=5$ rays; the complex gain $\alpha_{il}$ and $\beta_{il}$ follow the complex Gaussian distribution $\mathcal{CN}(0,10^{-0.1PL(d)})$, where $PL(d)=\varphi_a + 10 \varphi_b \log_{10} (d) + \xi (dB)$ with $\xi \sim \mathcal{N}\left(0, \sigma^{2}\right)$, $\varphi_a=72.0,\varphi_b=2.92,\sigma=8.7$dB \cite{akdeniz2014millimeter}. The auxiliary variables $\{t_{k,j}\}$ are initialized following $\mathcal{C}\mathcal{N}(0,1)$. The penalty factor is initialized by $\rho=10^{-3}$. Other system parameters are set as follows unless specified otherwise later: $K=3, F_2=6, d_{RIS}=50, c=0.9, \epsilon_1=10^{-4}, \epsilon_2=10^{-7}, \gamma_k=10$dB, $\sigma^2_k=-85$dBm, $\forall k \in \mathcal{K}$. All simulation curves are averaged over $100$ independent channel realizations.
\vspace{-0.1cm}
\subsection{Convergence Performance}
\begin{figure}[t]
\begin{centering}
\vspace{-0.5cm}
\includegraphics[scale=.5]{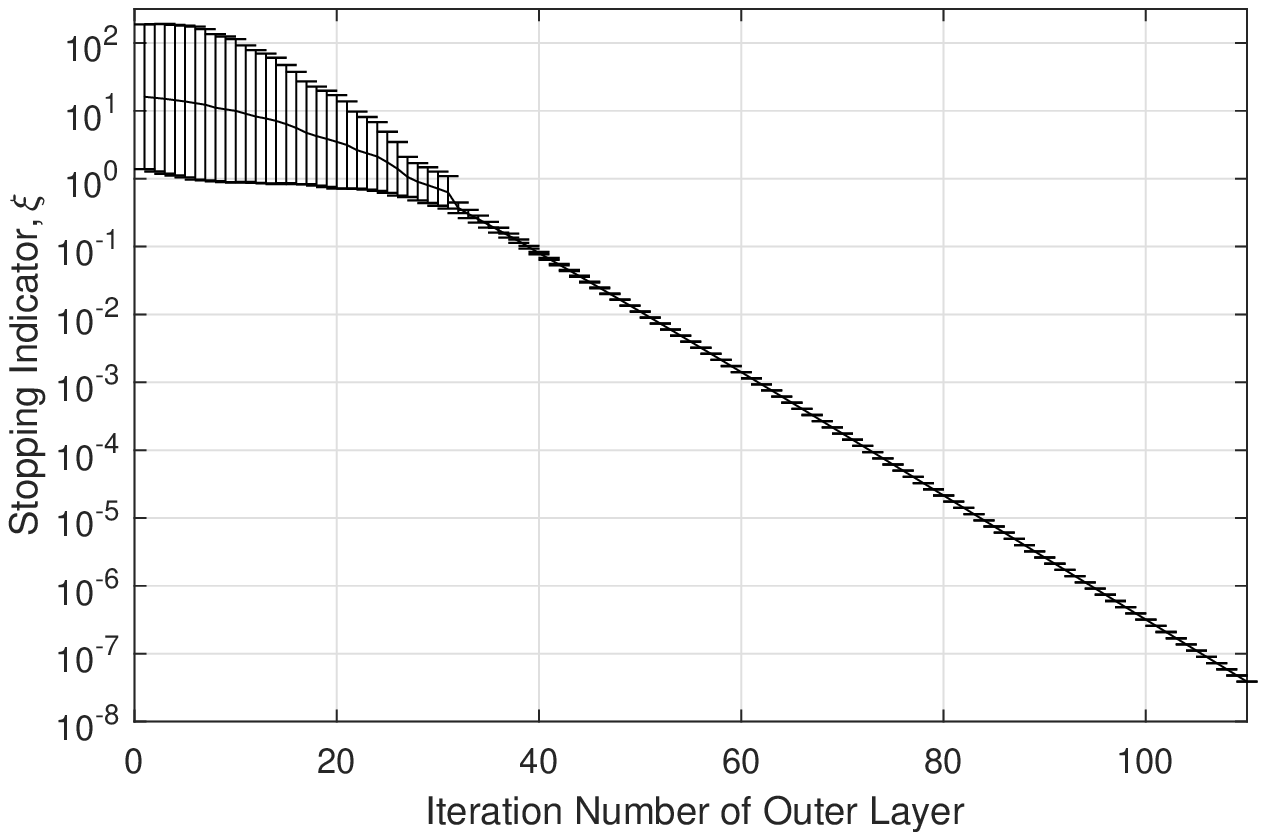}
\vspace{-0.1cm}
 \caption{\small{Stopping indicator of Penalty-based Algorithm}}\label{fig_conv_penalty1}
\end{centering}
\vspace{-0.5cm}
\end{figure}

\begin{figure}[t]
\begin{centering}
\vspace{-0.5cm}
\includegraphics[scale=.5]{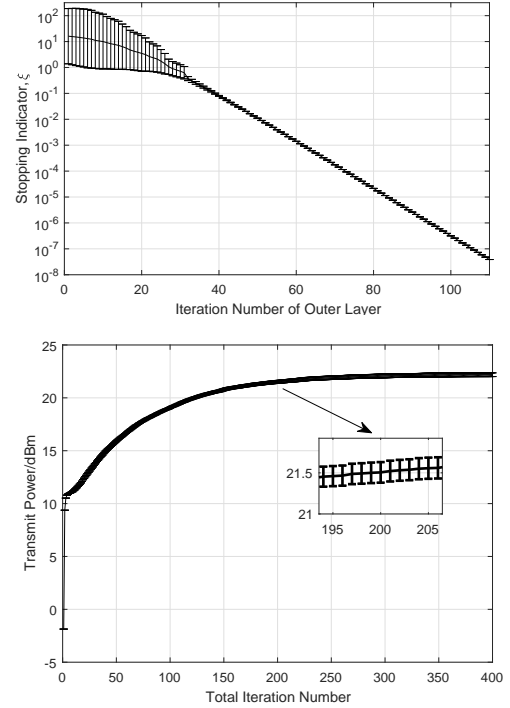}
\vspace{-0.2cm}
 \caption{\small{Convergence of Penalty-based Algorithm}}\label{fig_conv_penalty2}
\end{centering}
\vspace{-0.4cm}
\end{figure}

\vspace{-0.1cm}
We show the stopping indicator \eqref{stop_criteria} of the penalty-based algorithm in Fig.~\ref{fig_conv_penalty1} and the average convergence of the penalty-based algorithm in Fig.~\ref{fig_conv_penalty2} . These curves are plotted with the average plus and minus the standard deviation. It is observed that the stopping indicator can always meet the predefined accuracy $10^{-7}$ after about 110 outer layer iterations in Fig.~\ref{fig_conv_penalty1}. Thus, solutions obtained by the \textit{Algorithm} \ref{alg_penalty} satisfy all SINR constraints. Fig.~\ref{fig_conv_penalty2} shows that the proposed algorithm converges after about 300 total iteration numbers, which means that the inner layer runs averagely 3 times.

\begin{figure}[t]
\begin{centering}
\includegraphics[scale=.46]{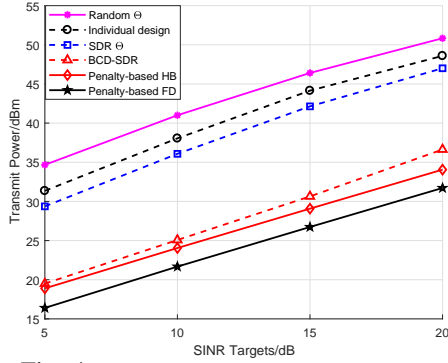}
\vspace{-0.25cm}
 \caption{\small{Transmit power versus SINR targets}}\label{fig_multi_sinr}
\end{centering}
\vspace{-0.5cm}
\end{figure}

\begin{figure}[t]
\begin{centering}
\vspace{-0.1cm}
\includegraphics[scale=.5]{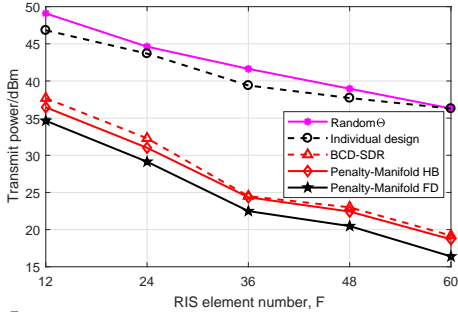}
\vspace{-0.25cm}
 \caption{\small{Transmit power versus the element number of RIS}} \label{fig_ele}
\end{centering}
\vspace{-0.4cm}
\end{figure}

\begin{figure}[t]
\begin{centering}
\vspace{-0.1cm}
\includegraphics[scale=.5]{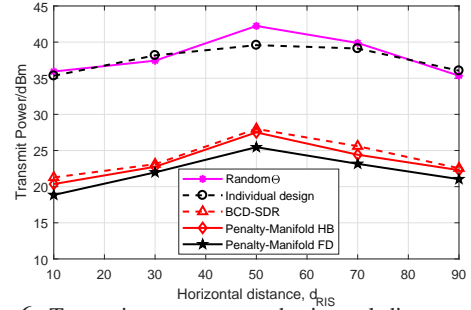}
\vspace{-0.3cm}
 \caption{\small{Transmit power versus horizontal distance of RIS}}\label{fig_distance}
\end{centering}
\vspace{-0.5cm}
\end{figure}

\subsection{Performance Comparison with other schemes}
To demonstrate the efficiency of the proposed algorithms and to reveal some design insights, we compare the performance of the following algorithms:
\begin{itemize}
\item Penalty-Manifold joint design with hybrid BF structure (Penalty-Manifold HB):  This is the proposed \textit{Algorithm} \ref{alg_penalty} for joint design of hybrid BF and RIS phase shifts.

\item Penalty-Manifold joint design with fully digital BF structure (Penalty-Manifold FD): This is the proposed \textit{Algorithm} \ref{alg_penalty} but changing the hybrid BF to fully digital BF at the BS. This is done by letting $D=1$.

\item Penalty-Manifold joint design with random $\mathbf{\Theta}$ (Random $\mathbf{\Theta}$):  The phase shifts at the RIS are randomly selected to be feasible values. Then the hybrid beamforming matrices $\{\mathbf{W},\mathbf{V}\}$ at the BS are obtained by using the penalty-manifold joint algorithm as in \textit{Algorithm} \ref{alg_penalty}, where the update of $\mathbf{\Theta}$ is skipped. This is to find out the significance of optimizing the phase shifts at the RIS.

\item Penalty-Manifold joint design with SDR $\mathbf{\Theta}$ (SDR $\mathbf{\Theta}$):  The phase shifts at the RIS are designed for maximizing the effective channel of the worse-cast user by using the SDR approach based on \eqref{max_min_ris}. Then the hybrid beamforming matrices $\{\mathbf{W}, \mathbf{V}\}$ at the BS are obtained by using the penalty-manifold joint algorithm as in \textit{Algorithm} \ref{alg_penalty}, where the udpate of $\mathbf{\Theta}$ is skipped. This is again to find out the significance of optimizing the phase shifts at the RIS.

\item BCD-SDR joint design (BCD-SDR):  The conventional BCD method in conjunction with the SDR method, as mentioned in the end of Section \ref{pro_formulation}.

\item  Individual design:  the proposed individual design where RIS phase shifts, analog BF, and digital BF are optimized sequentially in Section \ref{individual}.
\end{itemize}

Fig.~\ref{fig_multi_sinr} illustrates the transmit power versus SINR targets. We first observe that the Penalty-Manifold joint design outperforms the start-of-the-art BCD-SDR joint design. Second, the Penalty-Manifold joint design with random $\mathbf{\Theta}$ performs the worst among all the considered schemes. By simply changing the random $\mathbf{\Theta}$ to the SDR $\mathbf{\Theta}$ (while keeping the joint design of $\{\mathbf{W},\mathbf{V}\}$ unchanged), the transmit power consumption can be reduced by 5 dB. If $\mathbf{\Theta}$ is involved in the Penalty-Manifold joint design, another 10dB power reduction can be obtained. These observations indicate that the design of RIS phase shifts plays the crucial role for performance optimization.
Third, the individual design is about 2dB worse than the joint design with SDR $\mathbf{\Theta}$. This suggests that, when the RIS response matrix is designed individually for maximizing the effective channel gain of the worse-case user, further optimizing the hybrid BF at the BS can only bring marginal improvement.
Last but not least, the power consumed by Penalty-Manifold HB is about 2.5dB higher than the power consumed by Penalty-Manifold FD. Note that the hybrid BF has much lower hardware cost since it only employs $N=6$ RF chains, while the fully digital BF has $M=36$ RF chains.

The influence of the RIS element number, $F$, is considered in Fig.~\ref{fig_ele}. When $F$ increases from 12 to 60, the transmit power drops about 15dB. Thus, we conclude that the RIS can greatly reduce the transmit power by installing a large number of elements.

Fig.~\ref{fig_distance} illustrates the influence of the RIS location. It is seen  that as the RIS horizontal distance $d_{RIS}$ increases, the transmit power increases firstly, and reaches the peak at 50 m, then decreases. This can be explained that the received power through the reflection of the RIS in the far field is proportional to $\frac{1}{d_1^2 d_2^2}$, where $d_1$ and $d_2$ denote the distance between the BS-RIS and RIS-user, respectively. It is found that the RIS can be located near the BS or users to save energy.
\vspace{-0.2cm}
\section{Conclusion}
\vspace{-0.1cm}
This paper proposed a two layer penalty-based algorithm to solve the RIS-aided hybrid beamforming optimization problem in mmWave systems. In the inner layer, we alternately optimize the digital beamforming and analog beamforming at the BS and the response coefficient at the RIS. The outer layer updates the penalty factor to obtain a high precision solution. A low-complexity individual optimization method is also proposed. Extensive simulation results demonstrate that the proposed algorithm has a good performance and the RIS can significantly improve the energy efficiency. 

\bibliographystyle{IEEEtran}
\bibliography{hybrid2_1}

\end{document}